\documentclass[]{interact}
\usepackage[nice]{nicefrac}
\usepackage{anyfontsize}

\usepackage{color}
\usepackage{mathtools}
\usepackage{epstopdf}
\usepackage[caption=false]{subfig}

\usepackage[numbers,sort&compress,merge]{natbib}
\bibpunct[, ]{[}{]}{,}{n}{,}{,}
\renewcommand\bibfont{\fontsize{10}{12}\selectfont}

\theoremstyle{plain}

\theoremstyle{definition}

\theoremstyle{remark}

\begin{document}
\title{The Kramers turnover in terms of a macro-state projection on phase space}

\author{
\name{
Luca Donati\textsuperscript{a} 
\thanks{CONTACT Luca Donati. 
Email: donati@zib.de.
Address: Zuse Institute Berlin, Takustra\ss e 7, 14195 Berlin.
Telephone: +49 30 841 85 403.}
and 
Christof Sch\"utte\textsuperscript{b} 
and
Marcus Weber\textsuperscript{c}
}
\affil{
\textsuperscript{a,b,c}Zuse Institute Berlin, Takustra\ss e 7, 14195 Berlin;\\ 
\textsuperscript{a,b}Freie Universität Berlin, Department of Mathematics and Computer Science, Arnimallee 22, 14195 Berlin.
}
}

\maketitle

\begin{abstract}
We have investigated how Langevin dynamics is affected by the friction coefficient using the novel algorithm ISOKANN, which combines the transfer operator approach with modern machine learning techniques.
ISOKANN describes the dynamics in terms of an invariant subspace projection of the Koopman operator defined in the entire state space, avoiding approximations due to dimensionality reduction and discretization.
Our results are consistent with the Kramers turnover and show that in the low and moderate friction regimes, metastable macro-states and transition rates are defined in phase space, not only in position space.
\end{abstract}


%
\section{Introduction}
The dynamics of high-dimensional chemical systems can be modeled as one-dimensional Langevin dynamics governed by the stochastic differential equation
\begin{equation}
\begin{dcases}
\dot q_t  & =  \frac{p_t}{m} \\
\dot p_t  & =  - \frac{d }{d q} V(q) - \gamma p_t + \sqrt{2 k_B T \gamma m} \, \eta_t \, ,
\end{dcases}
\label{eq:langevin}
\end{equation}
where 
$q_t$ and 
$p_t$ denote the position and the momentum of the system at time $t$ on a one-dimensional relevant coordinate,
$V:\mathbb{R}\rightarrow \mathbb {R}$ is the potential of mean force acting on the system along the relevant coordinate,
and $m$ is the effective mass of the system, which represents how much inertia the system has along the relevant coordinate.
The system is coupled to a thermostat at temperature $T$, with Boltzmann constant $k_B$, through the friction coefficient $\gamma$ and the stochastic force $\eta_t$ defined as Gaussian white noise with $\langle \eta_t \rangle = 0$ and $\langle \eta_t, \, \eta_{t'} \rangle = \delta(t-t')$, where $\delta(t-t')$ is the Dirac delta function.
One possible field of application are molecular processes that exhibit metastability, where the energy function $V(q)$ presents minima and maxima with energy barriers that exceed the thermal energy $k_B T$.

In this context, a fundamental problem is the calculation of transition rates between potential minima, more precise, between macro-states of the system.
Indeed, transitions between the macro-states represent the most interesting biochemical processes in many applications, e.g. the folding of an amino acid chain or the binding/unbinding process between a receptor and a ligand.
However, they are rare events, i.e. they occur on very large time scales compared to reference time scales such as the oscillation times of hydrogen atoms.
Consequently, they are difficult to simulate and analyze, e.g. by means of Molecular Dynamics (MD) simulations.

Over the last century and a half, various theories and methods have been developed to solve this problem and find analytical solutions for calculating rates.
The first approximate formula of the problem dates back to Arrhenius, who derived in 1884 \cite{Arrhenius1889} the proportionality equation
\begin{eqnarray}
    k \propto e^{-\beta E_b} \, ,
\end{eqnarray}
where $k$ denotes the escape rate, $\beta = \nicefrac{1}{k_B T}$ is the inverse of the thermal energy, and $E_b$ is the height of the energy barrier, also known as activation energy of the reaction.
Later, further theories were developed that apply to different contexts of chemistry and physics.
Particularly noteworthy is the work of Kramers, who in 1940 \cite{kramers1940brownian} studied transition rates for one-dimensional systems driven by the Langevin equation and derived three formulas that apply to low, moderate and high friction regimes.
The three formulas well reproduce the so called Kramers turnover, a curve describing the transition rate as a function of $\gamma$:
the transition rate is linear with the coefficient $\gamma$ at low friction, then, having reached a plateau, the rate decays inversely to $\gamma$.
However, Kramers' theory remains incomplete in some aspects that were only later resolved.
For example, Langer derived a formula for multidimensional systems that operates in high friction regime \cite{langer1969}, Chandler derived a formula that takes into account non-Markovian effects \cite{chandler1978}, Mel’nikov and Meshkov have found an expression that improves the prediction in the transition from low to moderate friction \cite{Melnikov1986}, and, Pollak, Grabert and H\"anggi found a single expression that covers the entire friction range using a normal mode approach \cite{Pollak1989}.
These and other methods, which we do not mention for the sake of brevity, fall into the category of model-based methods, i.e based on the physical model of the system under investigation.

In this paper, we study the dependence of Langevin dynamics on the friction coefficient $\gamma$ using its representation in terms of the Koopman operator \cite{Koopman1931, Koopman1932}, which allows to transform the nonlinear problem defined in eq.~\ref{eq:langevin} into a linear problem.
The price of this is that the finite-dimensional dynamics in phase space is transformed into an infinite-dimensional problem in the space of observable functions \cite{Baxter_1996, Klus2018}.
For this reason, we seek invariant subspaces of the Koopman operator with finite dimensions.
We use the ISOKANN algorithm \cite{Rabben2020}, a data-driven method that identifies membership functions that constitute a basis of an invariant subspace of the Koopman operator preserving the Markovianity of the projected process.
The peculiarity of ISOKANN is that it does not require the identification and the discretization of reaction coordinates, instead, membership functions can be estimated on states of the full space by means of machine learning techniques such as neural networks, overcoming the problem of the curse of dimensionality.

Membership functions are a generalization of ordinary crisp sets and characterize the metastable macro-states of the system preserving the time scales of the micro system when projected onto the macro-states \cite{Deuflhard2004,Weber2006thesis,Weber2018}. 
Using ISOKANN, we estimated the phase space membership functions and calculated the rates, which represent transitions on the phase space.
In this way, we reproduced a rate curve as a function of friction, which is analogous to the Kramers turnover.
However, our results show that in low and moderate friction regimes, the rates in position space are an approximation due to the loss of Markovianity of the dynamics defined in eq.~\ref{eq:langevin}.
To obtain a correct representation of dynamics, even in low and moderate friction, it is necessary to take momenta into account.
Thus, with this work we intend to open up a new perspective in reaction rate theory.
This is possible through the use of increasingly advanced machine learning techniques, such as ISOKANN, which allows for the estimation of rates as functions of the entire phase space, preventing errors induced by discretization or dimensionality reduction.
%
\section{Theory}
We briefly introduce the operator theory that is needed to project Langevin dynamics onto macro-states \cite{Schuette1999b}.
\subsection{Transfer operator approach}
The dynamics of a stochastic process solution of the Langevin equation defined in eq.~\ref{eq:langevin} is equivalently described by the dynamics of the time-dependent probability density $\rho_t(x)$ solution of the partial differential equation
\begin{eqnarray}
  \frac{\partial \rho_t(x)}{\partial t}  
  =
  \mathcal{Q}^* \rho_t(x) \, ,
  \label{eq:inf1}
\end{eqnarray}
where the operator $\mathcal{Q}^*$ defines the Fokker-Planck equation, or forward Kolmogorov equation, and $x=(q,p)\in \Gamma \subset \mathbb{R}^2$ represents the state of the system in the phase space.
The solution of eq.~\ref{eq:inf1} is formally written as 
\begin{eqnarray}
    \rho_{t+\tau}(x) 
    &=&
    \exp \left(\mathcal{Q}^*\,\tau \right)\rho_t(x) \\
    &=&
    \mathcal{P}_{\tau}\rho_t(x)  \, ,
    \label{eq:Propagator}
\end{eqnarray}
where $\mathcal{P}_{\tau}$ denotes the propagator of probability densities with stationary density
\begin{eqnarray}
    \lim_{t \rightarrow +\infty} \rho_t(x) = \pi(x) \,,
\end{eqnarray}
defined by the Boltzmann distribution
\begin{eqnarray}
\pi(x) := \pi(q,p) = \frac{1}{Z} \exp\left(-\beta \left(\frac{p^2}{2m}  + V(q)\right)\right) \, ,
\label{eq:Boltzmann}
\end{eqnarray}
where $Z$ is a normalization constant.

Besides considering the evolution of probability densities, it is useful to study the evolution of observable functions $f(x)$.
To this end, we introduce the infinitesimal generator $\mathcal{Q}$, adjoint of the operator $\mathcal{Q}^*$ that defines the backward Kolmogorov equation
\begin{eqnarray}
  \frac{\partial f_t(x)}{\partial t} =
  \mathcal{Q} f_t(x) \, .
  \label{eq:Q}
\end{eqnarray}
Analogously to eq.~\ref{eq:Propagator}, we can write a formal solution of eq.~\ref{eq:Q} as
\begin{eqnarray}
    f_{t+\tau}(x) 
    &=& 
    \exp \left(\mathcal{Q}\,\tau \right)f_t(x) \\
    &=&
    \mathcal{K}_{\tau}f_t(x)  \\
    &=&
    \mathbb{E}\left[ f(x_{t+\tau})\vert x_t = x\right]
    \, ,
    \label{eq:Koopman}
\end{eqnarray}
where we introduced the Koopman operator $\mathcal{K}_{\tau}$ which propagates the expectation value of observable functions.
%
%
\subsection{Rates from membership functions}
Consider the $\tau$-dependent eigenvalues $\lambda_{\tau,i}$ and the associated eigenfunctions $\Psi_i$ of the Koopman operator $\mathcal{Q}$ such that
\begin{eqnarray}
    \mathcal{Q} \Psi_i = \lambda_{\tau,i} \Psi_i\, .
\end{eqnarray}
If the dynamics is ergodic and not periodic, then the first eigenfunction is constant $\Psi_1=1$ and it is associated to the non-degenerate eigenvalue $\lambda_{\tau,1}=1$.
In reversible dynamics, the subsequent $n_c$ dominant eigenfunctions $\Psi=\lbrace \Psi_2,...,\Psi_{n_c}\rbrace$, associated to sorted and negative eigenvalues $\lambda_{\tau,2}>\dots>\lambda_{\tau,nc}$, exhibit positive and negative values, which allows for the identification of metastable macro-states. 
In the non-reversible case, real-valued functions which span an invariant subspace of the Koopman operator can be applied instead of eigenfunctions. 
Each point in state space is represented by a vector which comprises of the values of these finitely many ($n_c$) functions. 
These points can be mapped into a $(n_c-1)$-simplex whose vertices represent the metastable states whereas the edges represent the transitions. 
The algorithm PCCA+ \cite{Deuflhard2004,Weber2006thesis}, by means of a linear transformation, transforms the simplex into a standard simplex, i.e. a simplex whose vertices are unit vectors.
Accordingly, the set of dominant eigenfunctions is transformed into a set of membership functions $\chi=\left( \chi_1,\chi_2,\dots,\chi_{n_c}\right)^\top$, with $\chi_i:\Gamma \rightarrow [0,1]$, $\forall i=1,2,\dots,n_c$, such that $\sum_i \chi_i = 1$.
Membership functions characterise the membership of a state $x$ in the macrostates of the system and by exploiting the linearity of the Koopman operator the exit rate from a macrostate is estimated as
\begin{eqnarray}
\kappa = - \frac{1}{\tau} \log(a_1) \left( 1 + \frac{a_2}{a_1 - 1}\right) \, ,
\label{eq:chi_exit_rate}
\end{eqnarray}
where $a_1$ and $a_2$ are obtained solving the linear regression problem
\begin{eqnarray}
    \min\limits_{a_1,a_2}\Vert \mathcal{K}_{\tau} \chi(x) - a_1 \chi(x) + a_2 \Vert .
    \label{eq:lin_reg}
\end{eqnarray}
For a complete discussion about the $\chi$-exit rates and the derivation of eq.~\ref{eq:chi_exit_rate}, we refer to \cite{weber2017fuzzy}.
%
%
\subsection{ISOKANN}
The calculation of rates according to eqs.~\ref{eq:chi_exit_rate} and \ref{eq:lin_reg} requires the membership function $\chi$ and the propagated membership function $\chi_t = \mathcal{K}_{\tau} \chi(x)$.
We use ISOKANN \cite{Rabben2020, sikorski2022learning}, an iterative algorithm which modifies the Von-Mises-Algorithm \cite{Mises1929} as iteration scheme
\begin{eqnarray}
    f_{k+1} &=& \frac{
    \mathcal{K}_{\tau} f_k}{\lVert \mathcal{K}_{\tau} f_k \rVert} \, ,
    \label{eq:Von_Mises}
\end{eqnarray}
where the initial function $f_0$ is an arbitrary function and $\Vert \cdot \Vert$ is the supreme norm.
As $k\rightarrow \infty$, eq.~\ref{eq:Von_Mises} converges to the first eigenfunction of the Koopman operator:
\begin{eqnarray}
    \lim_{k\rightarrow \infty } f_{k+1} = 
    \Psi_1 = 1 .
\end{eqnarray}
In fact, by applying $\mathcal{K}_{\tau}$ iteratively to a function $f$, one obtains the same result as applying the operator with lag time $\tau$ tending to infinity, i.e. a constant function.

Consider now a two-metastable system, as the model studied by Kramers, then the Koopman operator has two dominant eigenfunctions $\Psi_1$ and $\Psi_2$, and the membership functions are written as
\begin{eqnarray}
\begin{cases}
        \chi_1 = b_1 \Psi_1 + b_2 \Psi_2 \\
        \chi_2 = 1 - \chi_1
\end{cases}\, ,
\end{eqnarray}
with $b_1$ and $b_2$ appropriate coefficients.
We introduce a linear transformation $S$ to prevent the convergence of the Koopman operator to $\Psi_1=1$, and retrieve information regarding the  eigenfunction $\Psi_2$ such that $\Psi_1$ and $\Psi_2$ span an invariant subspace of the Koopman operator.
For this purpose, we choose as $S$ the shift-scale function
\begin{eqnarray}
    S\mathcal{K}_{\tau} f_k = \frac{\mathcal{K}_{\tau} f_k - \min\left(\mathcal{K}_{\tau} f_k\right)}{\max\left(\mathcal{K}_{\tau} f_k\right) - \min\left(\mathcal{K}_{\tau} f_k\right)} \, ,
    \label{eq:S}
\end{eqnarray}
that guarantees that $f_k:\Gamma \rightarrow [0,1]$.
The algorithm defined in eq.~\ref{eq:Von_Mises} is rewritten as
\begin{eqnarray}
    f_{k+1} &=& 
    S\mathcal{K}_{\tau} f_k\, ,
    \label{eq:Isokann}
\end{eqnarray}
and converges to one of the two membership function:
\begin{eqnarray}
    \lim_{k\rightarrow \infty } f_{k+1} &=& \chi_i \
    \quad i=1 \ \rm or \ 2 .
\end{eqnarray}
In general, we do not have an analytical representation of the Koopman operator or do not discretize the entire state space to retrieve its matrix representation.
However, we can calculate the action of the Koopman operator on observable functions applied to specific states in space $\Gamma$.
Exploiting the ergodic property, we approximate the expectation in eq.~\ref{eq:Koopman} as a time average:
\begin{eqnarray}
   f_{t+\tau}(x) 
    &\approx& \bar{f}(x_{\tau}) \\
    &=& \frac{1}{N} \sum_{n=1}^N f(x_{\tau,n}) \, ,
    \label{eq:Koopman_approx}
\end{eqnarray}
where $x_{\tau,n}$ are the final states of $N$ trajectories, solutions of eq.~\ref{eq:langevin}, starting at $x_0=x$.
Thus, eq.~\ref{eq:Isokann} is rewritten as
\begin{eqnarray}
    \bar{f}_{k+1}(x_0) &=& S\bar{f}_k(x_{\tau}) \, ,
    \label{eq:Isokann2}
\end{eqnarray}

Regarding the choice of the initial function $f_0$, a wide range of options is available.
The function should be an interpolating function that can be trained at each iteration until it converges to one of the membership function.
For higher dimensional systems, the use of neural networks is recommended, as was used in ref.~\cite{Rabben2020}.
However, for low-dimensional systems, other interpolation techniques may be used, e.g. spline functions or radial basis functions.
%

%
\section{Results}
\begin{figure}[ht!]
    \centering
    \includegraphics[width=0.5\textwidth]{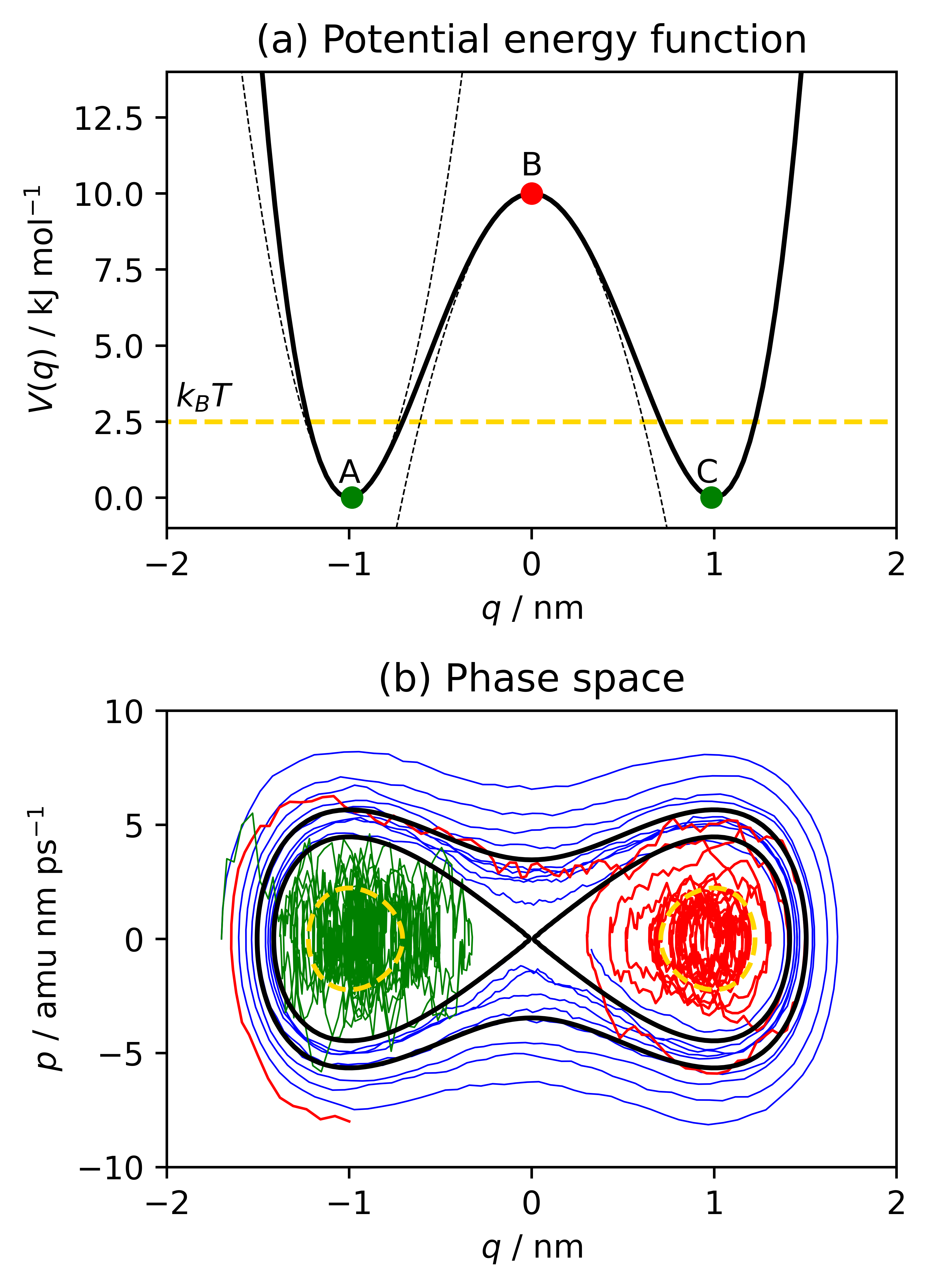}
    \caption{(a) Potential energy function (solid line) and harmonic approximation at the bottom of the wells and at the top of the barrier (dashed lines);
             (b) Phase space with energy levels (black contour lines and yellow dashed line denoting the $K_B T$ value) and three trajectories carried out with friction coefficients: 
             $\gamma = 0.1\,\mathrm{ps^{-1}}$ (blue),
             $\gamma = 2.2\,\mathrm{ps^{-1}}$ (red),
             $\gamma = 30.0\,\mathrm{ps^{-1}}$ (green).}
    \label{fig:1}
\end{figure}

As an illustrative example, we considered the Langevin dynamics of a fictitious particle of mass $m = 1\, \mathrm{amu}$ which moves in a one-dimensional potential energy function
\begin{eqnarray}
    V(q) 
    &=& 
   10 (q^2 - 1)^2 \ \mathrm{kJ\,mol^{-1}}\, .
    \label{eq:potential}
\end{eqnarray}
The function is characterized by two wells with minima at $q_A=-1\,\mathrm{nm}$ and $q_C=1\,\mathrm{nm}$, and a height barrier of $10\, \mathrm{kJ\,mol^{-1}}$ at $q_B=0\,\mathrm{nm}$ as illustrated in fig.~\ref{fig:1}-(a).
For our numerical experiments, we used standard thermodynamic parameters:
the temperature of the system was $T=300\, \mathrm{K}$ and the molar Boltzmann constant was $k_B = 8.314\times10^{-3} \,\mathrm{kJ\, K^{-1}\, mol^{-1}}$.
%
\subsubsection{Classic Kramers turnover}
In order to reproduce the classic Kramers turnover, we selected 25 friction coefficient values $\gamma$ between $0.1\,\mathrm{ps^{-1}}$ and $30.0\,\mathrm{ps^{-1}}$ and we solved the Langevin eq.~\ref{eq:langevin} using the Br\"unger, Brooks and Karplus (BBK) integrator scheme \cite{bbk1984} with an integrator timestep $\Delta t = 0.005\,\mathrm{ps}$.
For each value of $\gamma$, we ran 500 simulations starting at the bottom of the left well of the potential with an initial momentum randomly drawn from the Boltzmann distribution.
After the particle reached the bottom of the right well, the simulations were stopped and we calculated the mean time, i.e. the Mean First Passage Time (MFPT) $\langle \tau_{fp} \rangle$ \cite{pontryagin1933statistical}, from which we obtained the transition rate as
\begin{eqnarray}
    k_{A\rightarrow C} = \frac{1}{\langle \tau_{fp} \rangle} \, .
    \label{eq:mfpt}
\end{eqnarray}
The results of this numerical experiment are reported in fig.~\ref{fig:2}-(a) as black squares.
If the friction is very low ($\gamma \approx 0.1\,\mathrm{ps}^{-1}$), the dynamics of the system (eq.~\ref{eq:langevin}) is almost deterministic, and the system, unless it has enough initial momentum, is trapped in the well with an extremely low probability of escape. 
Correspondingly, the MFPT is very large and the value of the rate tends to zero.
However, increasing the friction by a small amount ($\gamma \approx 1.5\,\mathrm{ps}^{-1}$) the system gets enough thermal energy through the random force and increases the probability to escape from the well.
In fact, for low values of $\gamma$, the stochastic force $\sqrt{2k_BT\gamma m}\,\eta_t$, which is weighted by the square root of the friction coefficient, dominates the friction force $-\gamma p_t$, which is linear with the friction coefficient.
Thus, we observe a rapid and linear increase in rates up to a maximum of $k_{A\rightarrow C} \approx 0.02\,\mathrm{ps}^{-1}$.
Beyond the threshold of $\gamma \approx 1.5\,\mathrm{ps}^{-1}$, the system enters what is called the moderate friction regime.
Here, the friction force dominates the Langevin equation, and the probability of escaping the well, despite the high thermal energy, decays as $k_{A\rightarrow C} \propto \sqrt{1 + \nicefrac{1}{\gamma^2}}$.
For higher values of the friction coefficient ($\gamma>20\,\mathrm{ps}^{-1}$), the friction term is so strong that the average acceleration of the system tends to zero.
The dynamics is overdamped and the transition rate decays as $k_{A\rightarrow C} \propto \nicefrac{1}{\gamma}$.

The three friction regimes, here qualitatively described, were formalized by Kramers in 1940 \cite{kramers1940brownian}.
He assumed a two-metastable system governed by the Langevin dynamics with thermal energy $k_B T\ll E_b^+ = V(q_B) - V(q_A)$, so as to ensure metastability.
In addition, he required that the left well and the top of the barrier of the potential $V(q)$ are approximated by harmonic potentials with angular frequencies
\begin{eqnarray}
\omega_A = \sqrt{\frac{1}{m}\left.\frac{d^2 V}{dq^2 }\right\vert_{q_A}}\,,
\
\mathrm{and}\
\omega_B = \sqrt{\frac{1}{m}\left.\frac{d^2 V}{dq^2}\right\vert_{q_B}}\,.
\end{eqnarray}
Under these conditions, Kramers derived a transition rate formula for the low friction regime ($\gamma < \omega_B$)
\begin{equation}
    k_{A \rightarrow C} = 
    \frac{1}{2} \gamma \beta E_b^+
    \exp\left[-\beta E_b^+   \right]\, ,
    \label{eq:KramersWeakLimit}
\end{equation}
for the moderate friction regime ($\gamma > \omega_B$)
\begin{eqnarray}
k_{A \rightarrow C}
&=&
\frac{\gamma}{\omega_{B}} \left(\sqrt{\frac{1}{4} + \frac{\omega_{B}^2}{\gamma^2}} - \frac{1}{2} \right)
\cdot \frac{\omega_A}{2 \pi} 
\exp\left[-\beta E_b^+   \right] \, ,
\label{eq:KramersModerate}
\end{eqnarray}
and the high friction regime ($\gamma \gg \omega_B$)
\begin{eqnarray}
    k_{A \rightarrow C}
    &=& \frac{\omega_{B}}{\gamma}\cdot
    \frac{\omega_A}{2\pi}
    \exp\left[-\beta E_b^+   \right]
    \, .
    \label{eq:KramersHigh}
\end{eqnarray}
Note that Kramers defines the three regimes by comparing the coefficient of friction with the angular frequencies of the harmonic potentials that approximate the potential.
In fact, the transition probability also depends on the curvature near the pit and barrier.
The prediction of Kramers' formulas, reported in fig.~\ref{fig:2}-(a), is excellent, it is only around the threshold separating the low and moderate friction regimes that the model becomes inaccurate.
For more details about Kramers theory, we recommend Refs.~\cite{Hanggi1990,Peters2017}.
\subsubsection{Kramers turnover of membership functions in phase space}
In the second numerical experiment, we estimated the transition rates applying ISOKANN to the same setting of the first experiment.
We generated 1000 random initial points $x_0=(q_0,p_0)$ from a uniform distribution over the phase space defined by the $q$-range $[-2.0,\,2.0]$ nm and the $p$-range $[-10.0,\,10.0]$ amu nm ps$^{-1}$, and for each state we simulated $N=100$ trajectories of length $\tau = 7$ ps, corresponding to 1400 timesteps using a timestep integrator $\Delta t = 0.005$ ps.
The ISOKANN algorithm has been applied for 20 iterations using multiquadratic radial basis functions (RBF) \cite{2020SciPy-NMeth}, which are computationally undemanding and only require a few parameters to optimize during training, resulting in faster convergence.
We considered two cases:
\begin{itemize}
\item the membership functions $\chi_A(q)$ and $\chi_C(q)$, and the transition rate $k_{\chi_A\rightarrow\chi_C}$ between the macro-states of the position space; 
\item the membership functions $\hat{\chi}_A(q,p)$ and $\hat{\chi}_C(q,p)$ defined on the two-dimensional phase space, and the transition rate $\hat{k}_{\chi_A\rightarrow\chi_C}$ between macro-states of the phase space.
Note that from now on, each quantity marked with the symbol $\hat{\cdot}$ denotes a quantity measured over the entire phase space.
\end{itemize}
%
%
%

\begin{figure}[ht!]
    \centering
    \includegraphics[width=\textwidth]{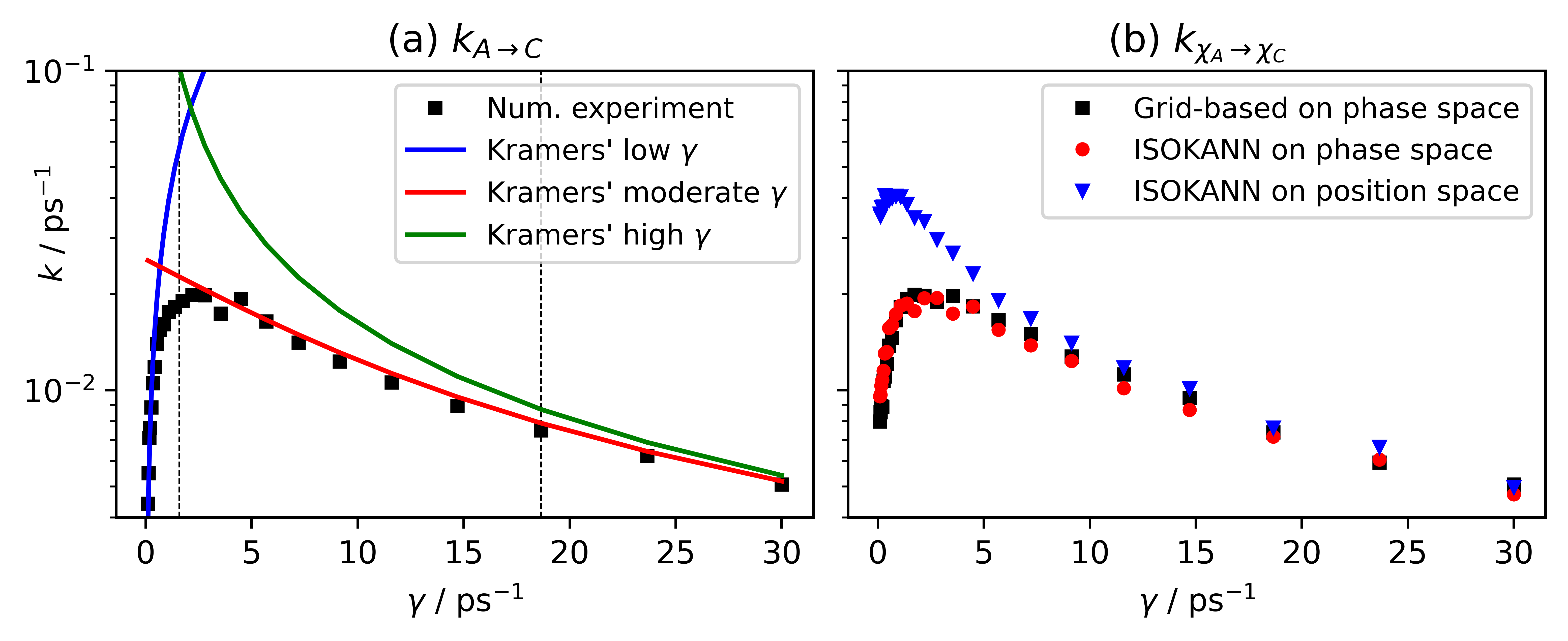}
    \caption{(a) Classic Kramers turnover: 
    transition rate $k_{A\rightarrow C}$ estimated by numerical experiment (black squares), 
    Kramers' formulas for low (blue), 
    moderate (red) and 
    high (green) friction coefficient $\gamma$. 
    The dashed vertical lines denote the threshold between friction regimes;
    (b) Kramers turnover between membership functions: 
    transition rate $\hat{k}_{\chi_A\rightarrow \chi_C}$ estimated by grid-based method (black squares) and
    ISOKANN (red circles),
    transition rate $k_{\chi_A\rightarrow \chi_C}$ estimated by ISOKANN (blue upside down triangles).}
    \label{fig:2} 
\end{figure}

\begin{figure}[ht!]
    \centering
    \includegraphics[width=\textwidth]{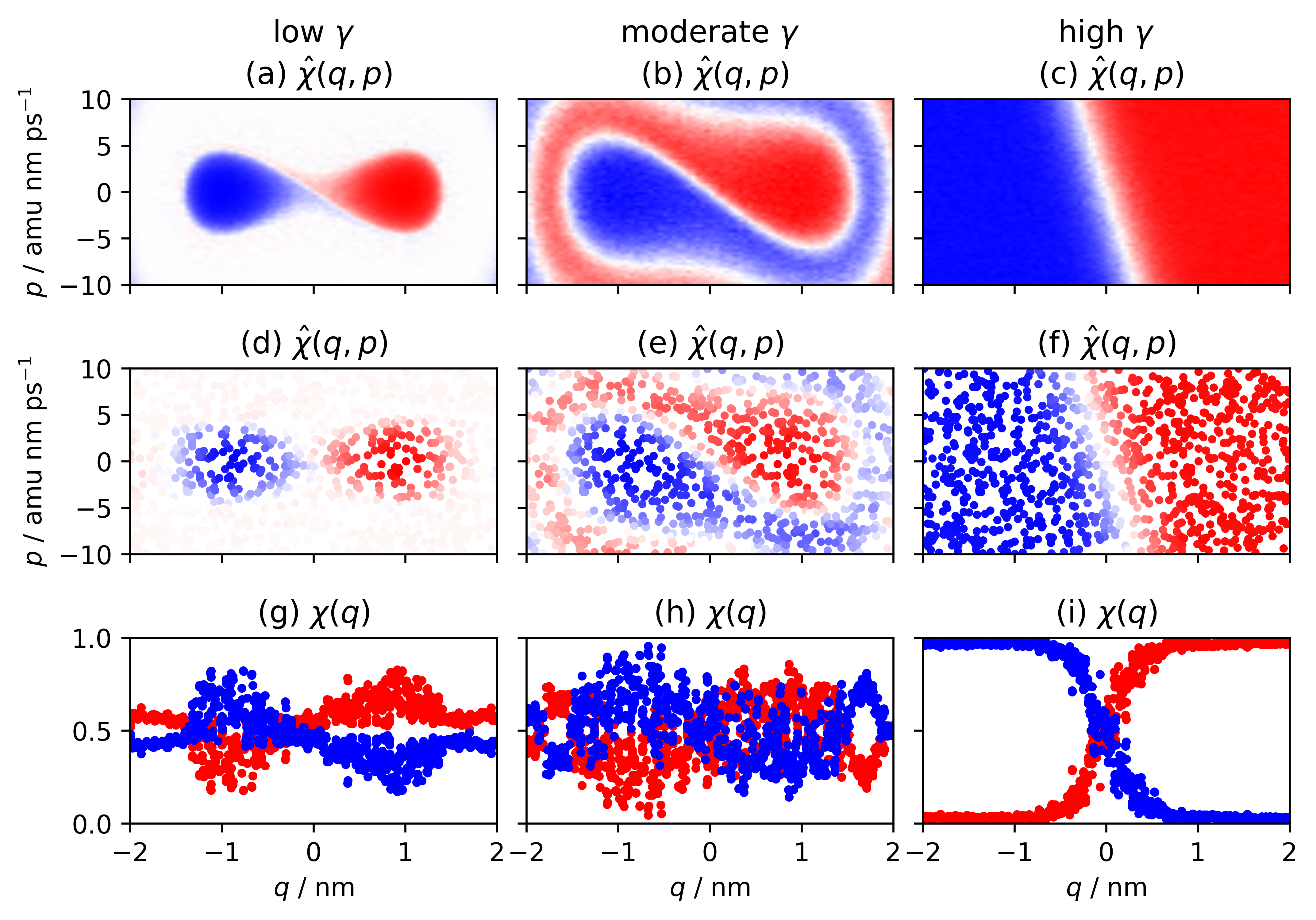}
    \caption{
    (a,b,c) Membership functions $\hat{\chi}(q,p)$ for $\gamma = 0.1,\,2.0,\,30\,\mathrm{ps^{-1}}$ estimated by grid-based model: 
    The blue-white-red colour gradient represents values in the range of 0 to 1.
    The membership functions are complimentary: $\chi_1+\chi_2=1$, then the blue points represent the macro-state $\chi_1$ and the red points represent $\chi_2$. The white points can be regarded as transitive regions.
    (d,e,f) Membership functions $\hat{\chi}(q,p)$ for $\gamma = 0.1,\,2.0,\,30\,\mathrm{ps^{-1}}$ estimated by ISOKANN;
    (g,h,i) Projection of the membership functions to position space.}
    \label{fig:3}
\end{figure}

The two rates, as functions of the friction coefficient $\gamma$, are reported in fig.~\ref{fig:2}-(b), respectively as blue upside down triangles and red circles.
Both curves show a turnover similar to the rate $k_{A\rightarrow C}$ reported in fig.~\ref{fig:2}-(a): rates have an ascending profile for very low range values, then, having reached the maximum ($k_{\chi_A\rightarrow\chi_C} \approx 0.4\,\mathrm{ps}^{-1}$ and $\hat{k}_{\chi_A\rightarrow\chi_C} \approx 0.2\,\mathrm{ps}^{-1}$), descend slowly.
However, while the values of the rate $\hat{k}_{\chi_A\rightarrow\chi_C}$ in phase space are overlapping with those of the Kramers rate $k_{A\rightarrow C}$ (although they are different physical quantities), the rate $k_{\chi_A\rightarrow\chi_C}$ defined in position space turns out to be higher in the low friction region but converges to the values of $\hat{k}_{\chi_A\rightarrow\chi_C}$ in the high friction regime.
To understand these results, it is useful to take a look at the membership functions obtained from ISOKANN and shown in fig.~\ref{fig:3}, where figures (d,e,f) on the second row and (g,h,i) on the third row are respectively the membership functions in the phase space and the position space, for low, moderate and high friction.
%
%
%
%
In fig.~\ref{fig:3}-(a) (low friction regime), the membership functions of the macro-states only have significant values for those states whose total energy $E=  \nicefrac{p^2}{2m} + V(q)$ is less than the height of the barrier.
The points with a total energy exceeding the barrier are depicted in white, indicating that they have an equal probability of belonging to either $\chi_A$ or $\chi_C$, approximately 0.5.
This occurs because trajectories originating from this area undergo periodic oscillations in phase space, visiting both wells as depicted in fig.~\ref{fig:1}-(b) by the blue trajectory.
Correspondingly, in fig.~\ref{fig:3}-(d), we show the membership function values as a projection of $\hat{\chi}_A(q,p)$ and $\hat{\chi}_C(q,p)$ onto the position space.
The apparent noise is due to the fact that the membership functions on phase space are not constant along the axis of momenta.
Therefore, when friction is low, the membership values projected to position space are not functions in a strict sense and do not describe position-based macro-states. 
In fig.~\ref{fig:3}-(b) (moderate friction regime), the membership functions draw concentric spirals that terminate in the minima of phase space respectively.
This may seem counterintuitive, but observing how a trajectory behaves in the moderate friction regime helps to interpret the membership functions correctly.
In fig.~\ref{fig:1}-(b), the red trajectory starts at position $q=-1\,\mathrm{nm}$ and momentum $p=-8\,\mathrm{amu\,nm\,ps^{-1}}$, and reaches the right-hand minimum by following a clockwise trajectory.
Similarly, if we start trajectories that are far from the central region of the phase space, we would observe spiral patterns that match with the membership functions.
From this figure, we deduce that in the moderate friction regime, the effective barrier, i.e. the transition region, is not at $q=0$, but between the two spirals.
In particular, in the central box $[-1,\,1]\times[-5,\,5]$ of the phase space, the barrier corresponds to a diagonal line which is not parallel to the momentum axis.
The reason is that if $q=0$ and $p>0$, the system reaches the right well with low probability of recrossing.
Conversely, if $q=0$ and $p<0$, the system reaches the left region.
Along the white diagonal the system is in an unstable equilibrium, i.e. the system has the same probability of reaching one of the wells, and ISOKANN assigns equal probability of membership to the two macro-states.
In fig.~\ref{fig:3}-(c) (high friction regime), membership functions are independent from momentum space.
The two regions $q<0$ and $q>0$ are assigned to the macro-states regardless of the momentum and the transition region is almost a vertical line.
The projections $\chi_A(q)$ and $\chi_C(q)$ onto the position space also appear well defined in fig.~\ref{fig:3}-(f).
This occurs because as the friction is very high, the momentum is quickly damped and it does not provide enough energy to overcome the barrier as shown by the green trajectory in fig.~\ref{fig:1}-(b). 
In the high friction regime, only the thermal noise can provide enough energy to jump over the barrier.
%
%
%
\subsubsection{Results validation}
In order to validate our results, we constructed a reference solution by means of a grid-based technique similar to Ulam's method \cite{ulam1960} which allows to discretize the operator $\mathcal{K}_{\tau}$ defined in eq.~\ref{eq:Koopman} into a transition probability matrix $\mathbf{K}(\tau)$, cf. \cite{Schütte_Klus_Hartmann_2023}.
Given a discretization of the phase space $\Gamma$ into $M$ disjoint subsets $\Gamma_i$, with $i =1,\dots,M$, and a set of $N$ simulations of length $\tau$ started in a random position of the subset $\Gamma_i$, then the entries of the matrix $\mathbf{K}(\tau)$ are written as
\begin{eqnarray}
    k_{ij}(\tau) = \frac{1}{N} \sum_{n=1}^N \mathbf{1}_{\Gamma_j} (x_{i,n}^{\tau})
\end{eqnarray}
where $\mathbf{1}_{\Gamma_j}$ is the indicator function
\begin{eqnarray}
\mathbf{1}_{\Gamma_j} (x) =
    \begin{dcases}
    1 ~&\text{ if }~ x \in \Gamma_j~, \\
    0 ~&\text{ if }~ x \notin \Gamma_j~\, ,
    \end{dcases}
\end{eqnarray}
and $x_{i,n}^{\tau}$ is the final state of the $n$th simulation started in $\Gamma_i$.
In practice, one counts how many times a simulation starting in $\Gamma_i$ ends in $\Gamma_j$ and divides by the number of simulations to obtain an estimation of the transition probability.
Afterward, an approximation of the infinitesimal generator, sometimes referred to as pseudogenerator, is obtained as
\begin{eqnarray}
    \widetilde{\mathbf{Q}} = \frac{\mathbf{K}(\tau) - \mathbf{I}}{\tau} \, ,
\end{eqnarray}
where $\mathbf{I}$ denotes an identity matrix of the same size as $\mathbf{K}{\tau}$.
Then the membership functions $\hat{\chi}(q,p)$ are calculated applying PCCA+ to $\widetilde{\mathbf{Q}}$ and the coarse-grained rate matrix between macro-states is calculated as a Galerkin projection of $\widetilde{\mathbf{Q}}$ onto the membership functions:
\begin{eqnarray}
\widetilde{\mathbf{Q}}_c
&=&
(\chi^\top \mathrm{diag}(\pi) \chi)^{-1} \chi^\top \mathrm{diag}(\pi) \widetilde{\mathbf{Q}} \chi  \, 
\label{eq:Qc}
\end{eqnarray}
In eq.~\ref{eq:Qc}, $\mathrm{diag}(\pi)$ denotes an $M\times M$ diagonal matrix, whose diagonal entries are the entries of the Boltzmann distribution $\pi(q,p)$ (eq.~\ref{eq:Boltzmann}) evaluated at the centers of subsets $\Gamma_i$.
Assuming a two-metastable system, the rate matrix $\widetilde{\mathbf{Q}}$ has size $2\times 2$:
\begin{eqnarray}
    \widetilde{\mathbf{Q}}_c = 
    \begin{pmatrix}
        -\widetilde{q}_{\chi_A\rightarrow\chi_C} & \widetilde{q}_{\chi_A\rightarrow\chi_C} \\
        \widetilde{q}_{\chi_C\rightarrow\chi_A} & -\widetilde{q}_{\chi_C\rightarrow\chi_A} \\
    \end{pmatrix} \, ,
    \label{eq:tildeQc}
\end{eqnarray}
with $\widetilde{q}_{\chi_A\rightarrow\chi_C} ,\, \widetilde{q}_{\chi_C\rightarrow\chi_A} > 0$ representing the transition rates between the macro-states.
For the sole case of a bimetastable system, these rates are equivalent to the exit rates defined in eq.~\ref{eq:chi_exit_rate}. 

Here, we discretized the $q$-range $[-2.0,\,2.0]$ nm in $80$ intervals of the same length $\Delta q = 0.05$ nm, and the $p$-range $[-10.0,\,10.0]$ nm in $70$ intervals of the same length $\Delta p = 0.29$ nm.
The transition rates estimated by PCCA+ are reported in fig.~\ref{fig:2}-(b) as black squares, while the membership functions are reported in fig.~\ref{fig:3}-(a,b,c).
For each subset, we ran 500 simulations of length 7 ps, with an integrator timestep of 0.005 ps for a total of 1400 timesteps.
There is excellent agreement between ISOKANN and the method based on the discretization of the phase space: both methods recreate the Kramers turnover and show the same patterns for the membership functions.
%
%
\section{Discussion and conclusion}
In this article, we studied the effect of the friction coefficient of Langevin dynamics on metastable macro-states of the phase space and calculated the transition rate.
%
%
%
%

For this purpose, we used the ISOKANN algorithm \cite{Rabben2020}, which identifies macro-states by means of membership functions that form a basis function of an invariant subspace of the Koopman operator.
In this subspace, the Koopman operator produces a linear dynamical system of finite dimensions that preserves the Markovianity of Langevin dynamics and can be used to determine kinetic observables such as transition rates.

We investigated a one-dimensional artificial potential, representing a bimetastable system, and reproduced the Kramers turnover.
However, differently from the original Kramers work, the transition rate we estimated represents transitions between macro-states in phase space.
Our results show that including both the positions and the momenta in defining the macro-states is necessary.
Indeed, neglecting the momentum in the low and moderate friction regime introduces non-Markovian effects that are not properly captured by the position-dependent membership functions.
In contrast, in the high friction regime, the velocity is instantaneously damped, and the macro-states can be defined as functions of only the position space.
%

%
%

This approach to estimating transition rates can be extended to highly dimensional problems.
The typical strategy requires solving the fundamental equations of motion, projecting the dynamics on a small number of relevant coordinates, and discretizing the low-dimensional model to create a matrix representation of the Koopman operator, as is done with Markov State Models \cite{Bowman2014,Wang2018b,Husic2018,keller2018} or Square Root Approximation of the infinitesimal generator \cite{Donati2018b, Donati2021, Donati2022b}.
The price of these techniques is the introduction of assumptions, such as the Markovianity of the projected dynamics, that can lead to significant errors \cite{Prinz2011}.
In contrast, ISOKANN does not require dimensionality reduction or space discretization, and the measured rates can be considered the best representation of the system's dynamics, net of approximations introduced a priori, e.g., when the equations of motion are numerically solved.
Thus, the dimensionality of the system poses no limits to ISOKANN on a theoretical level.
However, the implementation of ISOKANN for studying high-dimensional systems is more involved.
Here, considering the low-dimensionality of the system, we used radial basis functions, but for higher dimensional systems, we suggest more advanced interpolating functions such as feed-forward neural networks, which allow the use of all system coordinates, including momenta.
Another aspect to be taken into account is the choice of the mathematical representation of the molecular system.
Indeed, neural networks are not invariant with respect to translations and rotations when Cartesian coordinates are used as input data.
Thus, Cartesian coordinates must be transformed to a suitable set of input coordinates, for example pairwise distances, internal coordinates or atom-centred symmetry functions \cite{Behler2011}.

In summary, with this work, we have shown that ISOKANN is a valid tool for the study of dynamical systems that avoids the subspace projection of transfer operators.
Here, we have focused on the classic Kramers problem, studying how macro-states are defined in phase space and highlighting the importance of considering momenta in rate calculation.
Nevertheless, ISOKANN's flexibility and modern machine learning techniques allow for the study of even more complex systems.
\\
\\
\textbf{Acknowledgments} \\
\\
This research has been funded by the Deutsche Forschungsgemeinschaft (DFG, German Research Foundation) through the Cluster of Excellence MATH+, project AA1-15: ``Math-powered drug-design'' and the Collaborative Research Center CRC 1114, Projects B03 and B05.
\bibliographystyle{tfo}
\bibliography{references}

\end{document}